# Scattering theory and cancellation of gravity-flexural waves of floating plates

M. Farhat[1,*], P.-Y. Chen[2], H. Bagci[1], K. N. Salama[1], A. Alù[3], and S. Guenneau[4]

[1]*Division of Computer, Electrical, and Mathematical Science and Engineering, King Abdullah University of Science and Technology (KAUST), Thuwal 23955-6900, Saudi Arabia.*

[2]*Department of Electrical and Computer Engineering, University of Illinois at Chicago, Chicago, Illinois 60607, USA.*

[3]*Photonics Initiative, Advanced Science Research Center, City University of New York, New York, NY 10031, USA.*

[4]*Aix Marseille Univ, CNRS, Centrale Marseille, Institut Fresnel, Marseille, France.*

[*]mohamed.farhat@kaust.edu.sa

*We combine theories of scattering for linearized water waves and flexural waves in thin plates to characterize and achieve control of water wave scattering using floating plates. This requires manipulating a sixth-order partial differential equation with appropriate boundary conditions of the velocity potential. Making use of multipole expansions, we reduce the scattering problem to a linear algebraic system. The response of a floating plate in the quasistatic limit simplifies, considering a distinct behavior for water and flexural waves. Unlike similar studies in electromagnetics and acoustics, scattering of gravity-flexural waves is*

*dominated by the zeroth-order multipole term and this results in non-vanishing scattering cross-section also in the zero-frequency limit. Potential applications lie in floating structures manipulating ocean waves.*

In recent years, there has been a growing interest in studying the scattering of various kinds of waves from random and composite media [1]-[2]. Scattering cancellation is an active topic of research directly related to scattering analysis, and it relies on coating objects with shells of opposite dipole moment to cancel their scattering response in the quasistatic limit [3-5]. This technique has been generalized to account for various types of waves [2]. In the same vein, the practical importance of designing offshore floating structures and buildings, such as airports or oil plants, triggered the interest in characterizing water wave propagation and its effects on these structures [6-18]. These structures can be modeled as thin plates, and their interaction with water waves obeys the biharmonic equation [19]. In this Letter, we develop a scattering theory to rigorously formulate this interaction. This theory can have interesting applications in protection of harbors or offshore platforms from destructive waves, such as tsunamis. Moreover, the present study is motivated by earlier work on manipulation of water waves via lensing [20] and cloaking [2] with structures clamped to the waterbed, which could be transposed to floating structures [21-23].

In this Letter, an equation with sixth order derivatives with appropriate boundary conditions is proposed to model the scattering of linearized water waves. It is derived from the combined Navier-Stokes and plate theories [24-28] and used in characterizing scattering from an object floating on shallow water as schematized in Fig. 1(a).

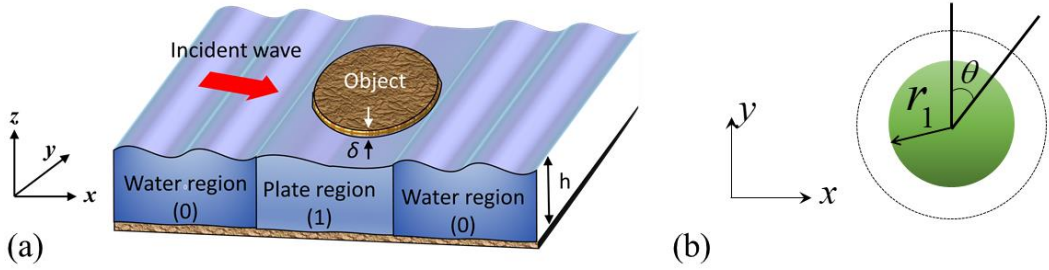

Figure 1 – (Color online) (a) A thin cylindrical plate of thickness $\delta$ floating atop water of depth $h$ and regions of the scattering problem. The water region is denoted by 0 and plate region by 1. (b) Top view of a cylindrical thin scatter to be considered in the first part of this study.

We then analyze the scattering response of the floating platonic structure shown in Fig. 1(b), which consists of a cylindrical disc in a thin plate, in the presence of a water wave excitation (time-harmonic vibration of the liquid free surface in the vertical $z$-direction). It is assumed that the out-of-plane dimension (the $xz$-plane in Fig. 1(a)) of the floating plate is negligible compared to its in-plane dimensions (the $xy$-plane in Fig. 1(a)) [24]. We show that in the quasistatic limit, i.e. for $\beta_0 r_1 \ll 1$, where $\beta_0$ is the bending wavenumber and $r_1$ is the size of the scatterer, the scattering is dominated by the zeroth order multipole term unlike the electrodynamics where the first significant order is the dipolar one. This is not the only marked difference between the two scenarios: the sixth order gravity-flexural partial differential equation (PDE), which typically describes the propagation of bending waves in ultra-thin plates floating atop incompressible fluids, is not equivalent to the vector/scalar wave equations that describe electromagnetic or acoustic wave propagation. Consequently, in view of Ref. [29], one can anticipate new types of Mie resonant modes and different wave physics.

The remainder of the Letter is organized as follows: We first formulate the problem. We then look for solutions in terms of multipole expansions. After that, we derive some asymptotic and numerical solutions to the scattering problem. We finally analyze the possibility of scattering cancellation in the context of gravity-flexural waves before giving some concluding remarks.

In the case of isotropic and uniform physical parameters, the equation governing gravity-flexural waves (in terms of the velocity potential) can be simplified to (See Eqs. (S1)-(S9) in Supplemental Material (SM) [30])

$$\Delta^3 \xi_1 + \beta_1^6 \xi_1 = 0, \tag{1}$$

where $\Delta$ is the Laplacian operator (See SM [30]) and $\beta_1 = (\rho \omega^2 / hD)^{1/6}$ is the gravity-flexural wavenumber, with $\rho$ the density of the fluid (taken to be 1000, except if otherwise stated) and $h$ its height (taken to be 10 m, except if otherwise stated). It should be understood that the operator is defined in the two-dimensional space. The wavenumber of the water wave is given by $k_0 = \omega / \sqrt{gh}$, in the case of shallow water approximation ($k_0 h \ll 1$). Otherwise, it has to satisfy the modified dispersion, i.e. $\omega^2 = gk_0 \tanh(hk_0)$. The dispersion relations for $\beta_1$ and $k_0$ are depicted in Fig. S1 [30] along with that for the flexural wavenumber $\beta = (M\omega^2 / D)^{1/4}$, where $M$ is the mass density of the plate. Similarly, for the water waves, one obtains in frequency domain

$$\Delta \xi_0 + k_0^2 \xi_0 = 0, \tag{2}$$

for the velocity potential in region 0. The reduced form of the sixth order PDE in velocity potential given in Eq. (1) governs the propagation of gravity-flexural waves. This PDE is supplemented with six boundary (continuity) conditions, in the case of a plate-plate interface. In the most general case of scattering these are namely $\xi$, $\partial_r \xi$, $\Delta \xi$, $\partial_r(\Delta \xi)$, $M_r(\Delta \xi)$, and $V_r(\Delta \xi)$, which expressions are given in the SM [30].

An object of radius $r_1$ is located atop an incompressible liquid (water in this study). For $r \leq r_1$ (inside the object), the gravity-flexural wavenumber is $\beta_1$. For $r > r_1$ (in region 0), the water wavenumber is $k_0$. The object is a thin plate that has the flexural rigidity $D(r) = D_1$, the Young modulus $E_1 = 10^8$ Pa, and the Poisson's ratio $v_1 = 0.25$. We verify that with our parameters, the thin plate approximation is always enforced (that is lateral dimension and gravity-flexural wavelengths are much higher than the plate's thickness). Without loss of generality, we assume that the object is illuminated by a water plane wave propagating in the $x$-direction, so that $kx = k_0 r \cos\theta$. The water wave velocity potential due to the incident plane wave is thus expressed as $\xi^{\text{inc}} = e^{ik_0 r \cos\theta}$, equivalently it can be expanded as

$$\xi^{\text{inc}}(r,\theta) = \sum_{n=0}^{\infty} \varepsilon_n i^n J_n(k_0 r) \cos n\theta,$$ where the coefficients $\varepsilon_0 = 1$ and $\varepsilon_n = 2$, $n \geq 2$.

At this point a re-writing of Eq. (1) is mandatory in order to expand the remaining displacement fields in terms of Bessel and Hankel functions of various kinds and orders. In fact, the velocity potentials $\xi(r,\theta)$ must be finite at $r = 0$ and satisfy the radiation condition at $r \to \infty$. Thus, Eq. (1) is recast as

$$\Delta^3 \xi_1 + \beta_1^6 \xi_1 = \left(\Delta + \beta_1^2\right)\left(\Delta - \beta_1^2\left(\frac{1}{2} + i\frac{\sqrt{3}}{2}\right)\right)\left(\Delta - \beta_1^2\left(\frac{1}{2} - i\frac{\sqrt{3}}{2}\right)\right)\xi_1$$
$$= \left(\Delta + \beta_1^2\right)\left(\Delta - \beta_1^2 \alpha\right)\left(\Delta - \beta_1^2 \alpha^*\right)\xi_1 = 0, \tag{3}$$

where we note $\alpha = 1/2(1+i\sqrt{3})$ to simplify the notation. And hence the displacement field $\xi_1$ is a superposition of solutions to the Helmholtz equation with real (first) and complex (second and third terms) conjugate gravity-flexural wavenumbers $\beta_\alpha = \beta_1 \sqrt{\alpha}$ and $\beta_\alpha^* = \beta_1 \sqrt{\alpha^*}$ with $\sqrt{\alpha} = 1/2(\sqrt{3}+i)$. So, the first term in Eq. (3) results in expansion in terms of $J_n(\beta_1 r)$. The second and third terms result in expansions in terms of $I_n(\beta_\alpha r)$ and $I_n(\beta_\alpha^* r)$, respectively, the modified Bessel functions of order $n$. Using all these assumptions, the field inside the thin-plate region is given by

$$\xi_1(r,\theta) = \sum_{n=0}^{\infty} \varepsilon_n i^n [B_n J_n(\beta_1 r) + C_n I_n(\beta_\alpha r) + E_n I_n(\beta_\alpha^* r)]\cos n\theta, \; r < r_1, \tag{4}$$

whereas the scattered elevation field in region 0, by taking into account Eq. (2) is

$$\xi_0^{\text{sca}}(r,\theta) = \sum_{n=0}^{\infty} \varepsilon_n i^n [A_n H_n^{(1)}(k_0 r)]\cos n\theta, \; r > r_1. \tag{5}$$

Here, $H_n^{(1)}(.)$, $J_n(.)$ and $I_n(.)$ are cylindrical Hankel functions of the first kind, Bessel and modified Bessel functions, respectively. To solve for the coefficients in the above equations, continuity relations (defined in the SM [30]) are used at the boundary at $r = r_1$, for each azimuthal order $n$. This yields a matrix system of equations in unknown coefficients $A_n$, $B_n$, $C_n$, and $E_n$ (the total size of the system is $4\times 4$, owing to the sixth order governing equation and number of boundaries).

The far-field scattering amplitude (or differential scattering cross-section) $f(\theta) = \sqrt{2r}e^{-i(k_0 r - \pi/4)} \lim_{r \to \infty} \xi_0^{sca}(r, \theta)$ is a measure of the object's visibility in direction $\theta$ and has the unit of a square root of length [26]. The total scattering cross-section, $\sigma^{sca}$, is the integral of $f(\theta)$ over all angles, i.e., $\sigma^{sca} = 1/2 \int_0^{2\pi} d\theta |f(\theta)|^2$. It may thus be expressed as

$$\sigma^{sca} = \frac{4}{k_0} \sum_{n=0}^{\infty} \varepsilon_n |A_n|^2 . \tag{6}$$

The unknown coefficients in Eqs. (4)-(5) satisfy the linear system $M_n X_n = X_{0,n}$

$$\begin{pmatrix} -H_n^{(1)}(k_0 r_1) & J_n(\beta_1 r_1) & I_n(\beta_\alpha r_1) & I_n(\beta_\alpha^* r_1) \\ -k_0 H_n^{(1)'}(k_0 r_1) & \beta_1 J_n'(\beta_1 r_1) & \beta_\alpha I_n'(\beta_\alpha r_1) & \beta_\alpha^* I_n'(\beta_\alpha^* r_1) \\ 0 & V_{J_n}(\beta_1 r_1) & V_{I_n}(\beta_\alpha r_1) & V_{I_n}(\beta_\alpha^* r_1) \\ 0 & W_{J_n}(\beta_1 r_1) & W_{I_n}(\beta_\alpha r_1) & W_{I_n}(\beta_\alpha^* r_1) \end{pmatrix} \begin{bmatrix} A_n \\ B_n \\ C_n \\ E_n \end{bmatrix} = \begin{bmatrix} J_n(k_0 r_1) \\ k_0 J_n'(k_0 r_1) \\ 0 \\ 0 \end{bmatrix},$$

(7)

with the expressions of the functionals $V_{Z_n}$ and $W_{Z_n}$ given in the SM [30]. The expression of $A_n$ can be obtained from Eq. (7) by using the Cramer's rule [24], i.e. $A_n = \det(M_{n,1}) / \det(M_n)$, where the matrix $M_{n,1}$ is obtained from $M_n$ by changing $H_n^{(1)}$ to $J_n$ (See Supplemental Materials for more details). If we consider very small scatterers, i.e., objects satisfying the quasistatic condition $k_0 r_1 \ll 1$ and $\beta_{1,\alpha} r_1 \ll 1$, only the first few of the scattering terms contribute to the scattering cross-section. In fact, solving Eq. (7) for $n = 0,1,2,3$ under the quasistatic condition yields the coefficients of these terms as

$$A_0 = \frac{2i\pi}{1 - 4\gamma_e + 2i\pi + \log(16) - 4\log(k_0 r_1)} + o\left((k_0 r_1)^{2/3}\right), \tag{8}$$

$$A_1 = -i\frac{3\pi}{16}(k_0 r_1)^2 + o\left((k_0 r_1)^2\right), \tag{9}$$

where $o(.)$ represents the Landau notation [24]. The expressions of $A_2$ and $A_3$ are given in the SM [30]. It can be seen that the dominant orders $A_0$ and $A_1$ are independent of the object's physical parameters and depend only of its geometrical dimension (radius). Also one has $A_{n\geq 2} = O\left((k_0 r_1)^{2(n+1)}\right)$ and $\rho_0 g / D_1$ relates the water wavenumber $k_0$ and gravity-flexural wavenumber $\beta_1$ through the relation $\beta_1^6 = (\rho_0 g / D_1) k_0^2$. This means that it has the unit of a length to the power $-4$, and thus the expression $r_1^4 (\rho_0 g / D_1)$ appearing in $A_2$ and $A_3$ is a dimensionless parameter. It can be seen that the dominating term is the zero-order coefficient $A_0$ unlike in electromagnetism where the dipolar scattering dominates. Additionally the observation that the contribution of $A_0$ to the scattering cross-section is infinite at zero frequency is intriguing, since it means that $\sigma^{sca}$ diverges at zero-frequency, which is counter-intuitive. The only example that possess somehow similar features is the example of pinned holes in thin-plates, in the context of flexural waves (obeying the biharmonic equation), as was analyzed in [26], [31], [28], [32]. However in that case, the scattering object was described only by its radius, i.e. there were no flexural field inside it, and the object has no other physical parameters, except its geometrical dimension. By contrast, in our study, the object is a thin plate, floating atop of water, and there is a field inside it, as can be seen from Eq. (4). The object has also physical parameters (density, Young modulus, Poisson's ratio, etc.). And still, the dominant scattering order $A_0$ is independent of all these properties, except the radius $r_1$. This observation if of tremendous importance to the last section of this Letter, where the scattering cancellation is considered for such class of waves. It is also reminiscent with zero-frequency bandgaps [31], [32]. Figure 2(a) gives the plot of the coefficients vs the normalized wavenumber and are compared against the numerical calculations using Eq. (7).

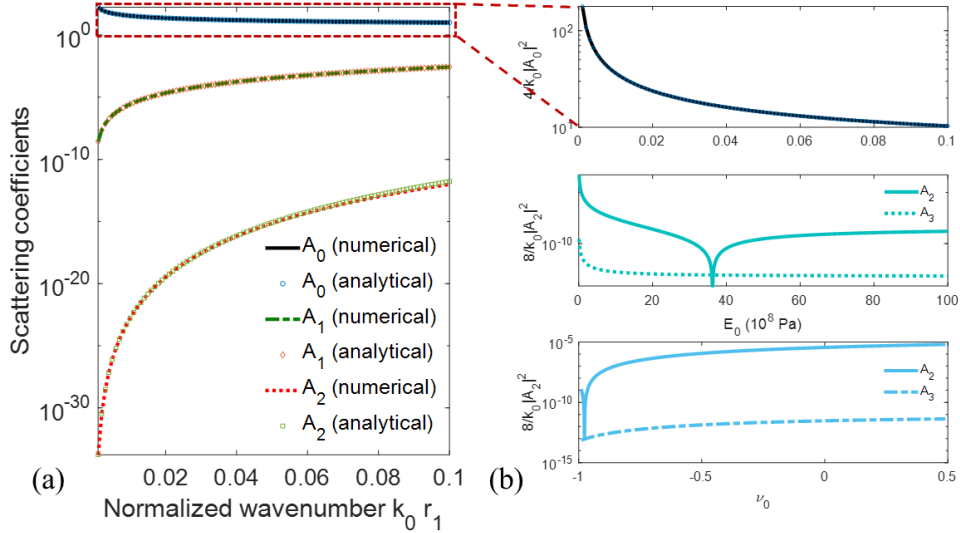

Figure 2 – (Color online) (a) Normalized scattering coefficients (for $n = 0, 1,$ and $2$) versus the normalized wavenumber obtained analytically, using Eqs. (8)-(9) and Eqs. (S15)-(S16) plotted using markers, and numerically using Eq. (7) plotted using lines. (b) Upper panel: zoom of the dominant scattering order $4/k_0 |A_0|^2$. Middle and lower panels give the dependence of $8/k_0 |A_2|^2$ and $8/k_0 |A_2|^2$ versus the Young modulus and Poisson's coefficient of the thin plate, normalized by $\iota_2$ and $\iota_4$, respectively. The vertical scale is logarithmic in all graphs, and note we consider a Poisson ratio $-1 < v_1 \leq 0.5$ that covers the case of auxetic plates.

The dependence of $A_n$, $n=2,3$ on the Young modulus and the Poisson ratio of the thin plate for the wavenumber $k_0 r_1 = 0.1$ are given in Fig. 2(b) middle and lower panels, respectively. These have dips (zeros) and they occur for $A_2$ when

$$r_1^4 (\rho_0 g / D_1)(303 - \nu_1(\nu_1 - 18)) = t_1(\nu_1 - 1)(\nu_1 + 3) \quad \text{and} \quad \text{for} \quad A_3 \quad \text{when}$$

$$r_1^4 (\rho_0 g / D_1)(19 + \nu_1)(3\nu_1 - 23) = t_3(\nu_1 - 1)(\nu_1 + 3).$$

The expressions given in Eqs. (8)-(9) and Eqs. (S15)-(S16) are only valid for small arguments. So to characterize the scattering from the objects shown in Fig. 1, one needs to numerically solve the algebraic system of Eq. (7) and compute the different scattering coefficients, and ultimately the scattering cross-section, and verify the convergence with respect to $N$, that is the number of coefficients used in Eq. (6). In this case, the dispersion relation of water waves shall be modified to the following relation $\omega^2 = g k_0 \tanh(h k_0)$. Then, by verifying the convergence of the scattering cross-section, we plot it versus the normalized wavenumber in Fig. 3(a), where it can be seen that multiple Mie resonances occur across the considered spectral region. This plot is given for a moderate value of the thickness of the plate, i.e. $\delta = 1$ cm, compared with the water wave wavelength. In this case for small wavenumbers, the quasistatic limit applies and the scattering coefficients follow Eqs. (8)-(9) and Eqs. (S15)-(S16). However, for very small thickness on the order of millimeter, this limit does not apply anymore, since $k_0 r_1 \ll 1$ but $\beta r_1 > 1$. The SCS is given for this scenario in Figs. 3(b)- 3(c), where the latter is a magnified view of the former. These plots show a completely different behavior, whereby one observes scattering maxima, for small wavenumbers, and these scattering resonances are dominant, compared to the classical higher frequency Mie resonance [33]. In Fig. 3(c) we can see that these resonances are ultra-narrowband and of Fano line-shape. We further note that the smaller the thickness of the plate, the higher the zero-frequency scattering, as can be clearly seen in Figs. 3(b)-(c).

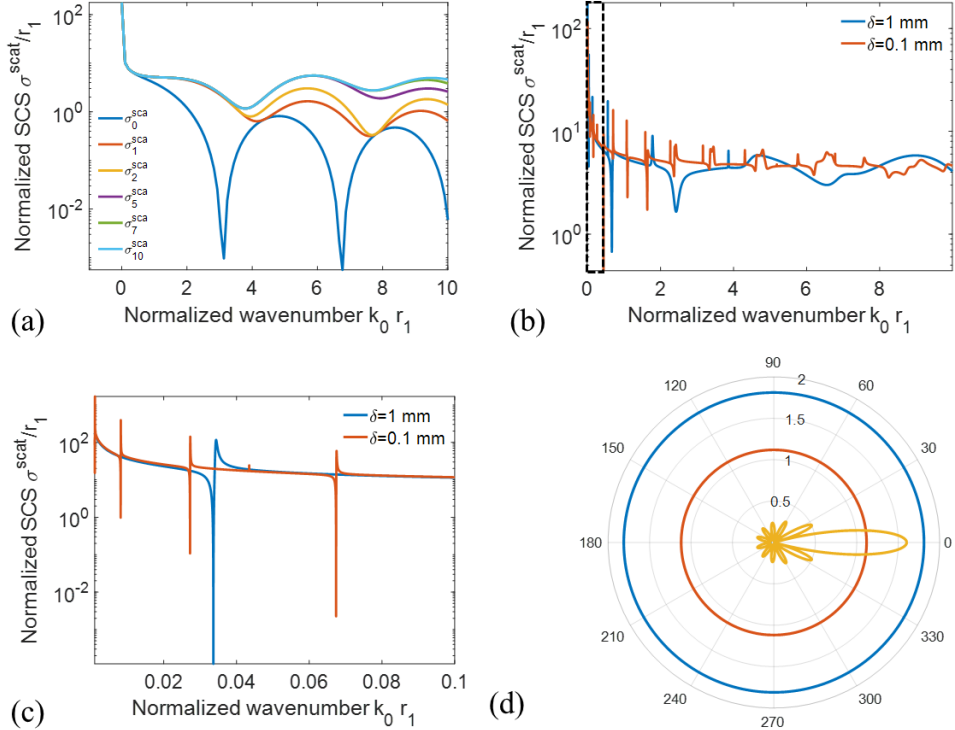

Figure 3 – (Color online) (a) Scattering cross-section of the cylindrical thin plate (with $\delta = 1$ cm, $v_1 = 0.25$, and $E_1 = 10^8$ Pa ), with varying number of multipole orders, until convergence is obtained. (b) Same as in (a) but for a thickness of the plate of 1 mm and 0.1 mm. (c) Magnified view of (b) in the low frequency regime. (d) Angular dependence of the scattering cross-section. Note the vertical axes in (b), (c) have a logarithmic scale.

The angular dependence of the SCS is depicted in Fig. 3(d), where we plot it versus the angle of observation for three different values of the wavenumber, i.e. $k_0 r_1 = 0.1$ (blue line), $k_0 r_1 = 1$ (blue line), and $k_0 r_1 = 10$ (yellow line). Enhanced and isotropic

Mie resonance can be seen, quite remarkably for the smaller wavenumber, in contrast to usual scattering scenarios. For the higher wavenumbers, the amplitude of the scattering is reduced and becomes strongly anisotropic. This features is unique to this kind of scattering. It should be mentioned too that for very small thickness [red curve in Fig. 3(c)] for some modes, the maximum is preceded by a minimum, where the SCS goes to zero, which is again reminiscent of Fano resonances. This also shows that the cylindrical plate becomes nearly invisible for some frequencies without coating. The analysis of cloaking and scattering cancellation for such waves will be analyzed in further works.

Let us now consider coating the object as can be seen in Fig. 4(a). An object of radius $r_1$ is thus coated with a shell of outer radius $r_2$, and both float atop an incompressible liquid (water in this study). For $r \leq r_1$ (inside the object, i.e. region 1), the gravity-flexural wavenumber is $\beta_1$. For $r_1 < r \leq r_2$ (inside the shell, i.e. region 2), the gravity-flexural wavenumber is $\beta_2$. And for $r > r_2$ (in region 0), the water wavenumber is $k_0$ as before. The object and the shell are thin-plates that have the flexural rigidity $D_{1,2}$, the relative Young modulus $E_{1,2}$, and the Poisson's ratio $v_{1,2}$, respectively. Without loss of generality, we assume that the core-shell structure is illuminated by a water plane wave propagating in the $x$-direction, so that $kx = k_0 r \cos\theta$. The water wave velocity potential is, as before $\xi^{inc}(r,\theta) = \sum_{n=0}^{\infty} \varepsilon_n i^n J_n(k_0 r) \cos n\theta$, where the coefficients $\varepsilon_0 = 1$ and $\varepsilon_n = 2$, $n \geq 2$. Using all these assumptions, the field inside region 1 is given by

$$\xi_1(r,\theta) = \sum_{n=0}^{\infty} \varepsilon_n i^n [B_n J_n(\beta_1 r) + C_n I_n(\beta_{\alpha,1} r) + E_n I_n(\beta_{\alpha,1}^* r)] \cos n\theta, \ r < r_1. \tag{10}$$

In region 2, i.e. inside the cloaking shell, one has

$$\begin{aligned}\xi_2(r,\theta) = \sum_{n=0}^{\infty} \varepsilon_n i^n [&F_n Y_n(\beta_2 r) + G_n K_n(\beta_{\alpha,1} r) + L_n K_n(\beta_{\alpha,2}^* r) \\ &+ N_n J_n(\beta_2 r) + O_n I_n(\beta_{\alpha,2} r) + P_n I_n(\beta_{\alpha,2}^* r)] \cos n\theta, \ r_1 < r \leq r_2,\end{aligned} \tag{11}$$

whereas the scattered velocity potential field (i.e. in region 0) is similar to Eq. (5). To solve for the 10 unknown coefficients in the above equations, continuity relations (defined in the SM [30]) are used at the boundary at $r = r_1$ (6 conditions) and at the boundary $r = r_2$ (4 conditions) for each azimuthal order $n$. This yields a matrix system of equations in the unknown coefficients (the total size of the system is $10 \times 10$, owing to the sixth order governing equation and the number of layers). Generally speaking, the possibility for an observer to detect the object in the far-field is determined by the value of $\sigma^{sca}$ given in Eq. (6). As a result, minimizing or completely canceling $\sigma^{sca}$ would lead to invisibility of the object in the far-field, irrespective of the observer's position. One question that arises at this stage is the possibility of cancelling the coefficients $A_n$ that significantly contribute to the scattering. The unknown coefficients in Eqs. (10)-(11) satisfy the linear system $M_n X_n = X_{n,0}$ (deduced from continuity conditions), with the matrix $M_n$ given in the SM [30]. The leading coefficients describing the scattering from the system are $A_n$. In order to obtain their expressions, one needs to use the Cramer's rule [24], i.e. $A_n = \det(M_{n,1})/\det(M_n)$, with matrices $M_{n,1}$ deduced from $M_n$ matrices by replacing their first column with the vectors $X_{n,0}$.

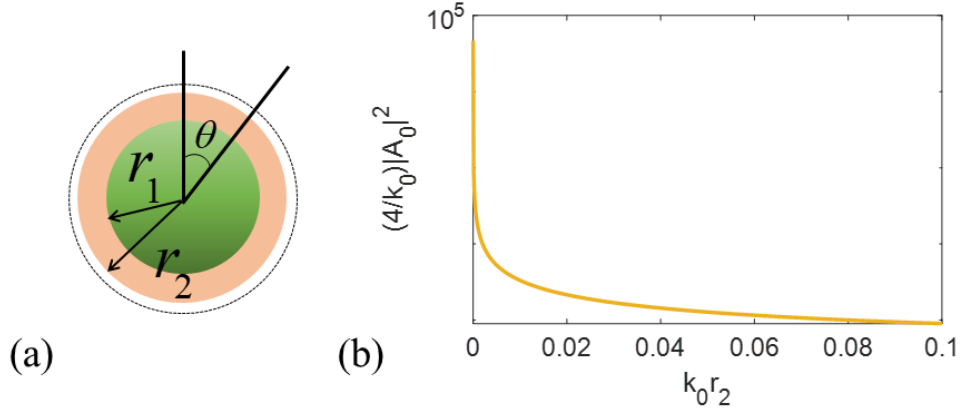

Figure 4 – (Color online) (a) Geometry of the core-shell structure. (b) Norm of the normalized scattering order of leading order, showing its divergence at zero frequency.

If we consider very small scatterers, i.e. objects satisfying the quasistatic condition $k_0 r_{1,2} \ll 1$ and $\beta_{\alpha,1} r_{1,2} \ll 1$, only the first few scattering coefficients terms $A_n$ contribute to the scattering cross-section. In fact, solving Eq. (S17) for $n = 0,1,2$, under the quasistatic condition yields

$$A_0 = \frac{2i\pi}{1 - 4\gamma_e + \log(16) - 4\log(k_0 r_2) + 2i\pi} + o\left((k_0 r_1)^{2/3}\right), \tag{12}$$

where $o(.)$ represents, as before the Landau notation and $\gamma_e$ is the Euler–Mascheroni constant which is approximately equal to 0.577215664901. Since $k_0 r_2 \ll 1$, the expression of Eq. (12) is a constant. Thus, it can be seen that irrespective of the physical parameters of the coating shell (i.e. flexural rigidity, Poisson's ratio, or density, etc.), $A_0$ cannot be made equal to zero, and thus scattering cancellation technique in this context is not possible. However, transformation coordinates technique can be used [23] to make such object invisible, but at the cost of complex physical parameters of the cloaking shell (non-uniformity and anisotropy). For very small frequencies, the denominator in Eq. (12) tends to infinity, but in logarithmic manner. However, if we normalize with the wavenumber, as can be seen from Eq. (12), one can see that the scattering becomes singular. In fact, one has

$$\frac{4}{k_0}|A_0|^2 \approx \frac{\pi^2}{\left(k_0 \log(k_0 r_2)\right)^2}. \tag{13}$$

And owing to the fact that $\lim_{x \to 0}(x \log x) = 0$, $\sigma^{sca}$ diverges in the zero-frequency limit. As can be seen further in Fig. 4 (b), this scenario of singular scattering is reminiscent of the case of clamped biharmonic (purely flexural) obstacles [26]. An analogous scenario occurs for microwaves of certain polarization in the case of a thin metal wire, and this has been used to dilute the average concentration of electrons and considerably enhance the effective electron mass through self-inductance [34]. It is interesting to note also that the dominant order scattering $A_0$ is independent of the parameters of the inner object, including its radius. This is interesting for potential applications in protection against long wavelength ocean waves as this indicates some zero frequency stop bands could be achieved for water waves propagating through an array of periodically distributed floating objects [32]. The analogous scenario occurring for Lamb waves propagating within periodically clamped plates has led to the design of an earthquake shield [34]. We further note zero frequency photonic stop bands associated with periodically distributed infinite conducting wires [35] are associated with non-commuting limits in homogenization theory, and a similar issue should arise for floating plates.

In this Letter, we analyzed in detail the scattering of gravity-flexural waves that propagate when an elastic thin plate lies atop a liquid incompressible surface (water for instance). These waves obey a sixth order partial differential equation, markedly different from the classical Helmholtz equation. Scattering from a single cylindrical objects was first investigated and low-frequency Mie resonances were shown to exist. Additionally, by coating the cylindrical object, scattering cancellation was shown to be impossible to realize, irrespective of the physical parameters of the shell, which is a paradigm shift compared with scattering cancellation for other types of waves. This unusual behavior can be understood, since the incident wave (gravity wave) is different from the waves that propagate inside the plate (gravity-flexural waves). Further studies are undergoing to investigate the scenario of a core shell structure with a gravity-flexural wave incident on it. Motivated by the search for zero frequency stop band structures in different wave systems, such as the recently achieved seismic shields in sedimentary soils structured by arrays of clamped columns to a bedrock [36], we would like to now build upon the present work to analyze Floquet-Bloch waves propagating within a doubly periodic array of floating plates. To do this, we make use of the Rayleigh method developed previously for fourth-order partial differential equations governing propagation of flexural waves in thin plates [31]. The Rayleigh method is also well suited for our sixth-order partial differential equation for gravity-flexural waves. We believe that our work opens unprecedented avenues in the control of water waves making use of floating metamaterial structures [32].